\documentclass[global,twocolumn,final]{svjour}
\usepackage{graphics}
\journalname{App. Phys. B}
\begin{document}
\title{Harmonic generation by atoms in circularly polarized laser fields: far-off and near resonances regimes}
\author{F. Ceccherini\inst{1}, N. Davini\inst{1}, D. Bauer\inst{2} \and F. Cornolti\inst{1}% etc
}                     % Do not remove
\institute{Istituto Nazionale per la Fisica della Materia (INFM), sez. A, Dipartimento di Fisica ``Enrico Fermi'', Universit\`a di Pisa, Via F. Buonarroti 2, 56127 Pisa, Italy \and Max-Born-Institut, Max-Born-Str.\ 2a, 12489 Berlin, Germany}
\date{27th October 2003}
\maketitle
\begin{abstract}
The generation of harmonics by atoms interacting with two circularly polarized and frequency related laser fields is addressed through {\em ab initio} numerical simulations. A detailed charaterization of a few specific harmonics is given.  In particular, the two different cases where the total energy  absorbed through photons is far-off or close to the energy gap between  different atomic states are investigated. It is found that the conversion efficiency in the harmonic generation  is strongly dependent on the inner atomic structure and in certain specific cases it can be significantly enhanced within a small frequency range.      
\end{abstract}
\section*{Introduction}  
\label{intro}
The interaction between an atom and two coplanar laser fields with circular polarization and commensurate frequencies is certainly an intriguing topic. In fact, it is worth to recall that when a single circularly polarized laser interacts with an atom no harmonic generation is possible.\\  
In the case where the two laser fields have opposite polarization the emitted harmonics are of order
\begin{equation}
\label{rules}
n = k(\eta +1) \pm 1,
\end{equation}
where $\eta$ is the frequency ratio and $k \in {\cal N}_+$. Selection rules of this type are characteristic of all systems whose Hamiltonian is invariant under a dynamical symmetry operation\cite{alon}, i.e., a transformation in both space and time. A further example of such systems is a circular molecule, e.g. benzene, interacting with a single circularly polarized laser field\cite{alon,vitali,cecch0,cecch1}. The derivation of such selection rules can be achieved by means of different arguments: angular momentum conservation, group theory and change of reference frame\cite{cecch1}. The configuration we investigate here  is of particular interest because of the filtering effect due to the selection rules (\ref{rules}), i.e., the higher is $\eta$ the less is the number of harmonics in a certain frequency range, and the circular polarization of the emitted harmonics. This last point becomes particular appealing for harmonics in the soft X-ray regime.  Harmonic generation from sources other than atoms and circular molecules, like linear molecules \cite{bandrauk}, nanotubes \cite{alonnanotube}, and plasmas \cite{linde} have been also investigated.
The present paper is organized as follows: after a short summary of the theoretical frame within which our system lies, a full description of the model and of the numerical simulations performed is presented. Finally, a conclusion is given. Atomic units (a.u.) are used throughout all the paper. \\

\section*{Theoretical frame}
\label{sec:1}
In order to fully understand the nature of the hamonics we investigate, let us firstly summarize the derivation of the selection rules (\ref{rules}) using angular momentum conservation.  Let $z$ be the propagation direction of the two lasers and $\sigma^+$, $\sigma^-$ the  polarizations of the laser of frequency $\omega$ and $\eta\,\omega$, respectively. If a harmonic $\sigma^+$ is emitted the sum of all the components along $z$ of the angular momentum carried by the absorbed photons has to be $+1$. If the atom for example absorbs $p$ photons from the low-frequency laser, it must absorb $p-1$ from the other laser. Hence, the total absorbed energy is $p\omega + (p-1)\eta \omega = \omega [p(1+\eta) - \eta] = \omega [(p-1)(1+\eta) +1]$. Since $p$ is arbitrary, with $k=p+1$ and $N=\eta +1$ we see that the harmonic of order $(kN+1)$ is emitted. With the same argument, starting from a harmonic of polarization $\sigma^-$, we  obtain $kN-1$. The key point that should be stressed is the following: harmonic generation is possible if, and only if, the absorption of $p$ photons from one laser is accompanied by the absorption of $p \pm 1$ photons from the other laser.\\  
Angular momentum considerations, as well as symmetry invariances,  provide informations on the allowed orders, but they cannot be used to derive any details concerning relative intensities and structures in the harmonic spectra like plateaus and cut-offs. Above all, stating that a certain order is allowed does not mean that the corresponding harmonic is really emitted. In fact, the previous arguments do not include case-specific features such as the atomic structure and the laser intensities. \\ 
Recently, the characteristics and the field dependence of the harmonics emitted in our configuration have been addressed analytically through a calculation\cite{cecch2} based on the Lewenstein model\cite{lewen}. With this analytical tool it is possible to derive a generalized cut-off for the harmonic spectra, an integral expression for the harmonic dipole strength, and to obtain again, independently, the selection rules (\ref{rules}). For what concerns the harmonic intensities such a model treatment holds as long as the total absorbed energy by the electron is not ``too close'' to a transition\cite{cecch2}. In case the absorbed energy is far from any resonance different predictions obtained from the analytical model are confirmed through the numerical simulations. In particular it was found that if $E_1$ and $E_2$ are the electric fields of the low- and high-frequency laser, the intensity of the two harmonics with the same $k$ are very different when $E_1 \ll E_2$, and they converge for $E_1 \gg E_2$.

\section*{Numerical simulations}
\begin{figure}
\resizebox{0.5\textwidth}{!}{%
  \includegraphics{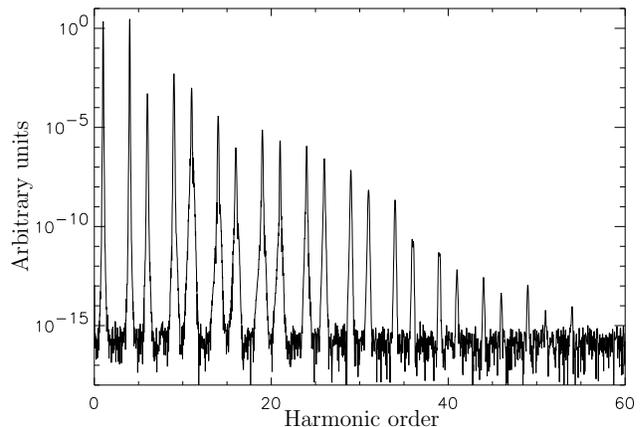}
}
\caption{Example of a typical harmonics spectrum obtained with two circularly polarized laser fields. The electric field intensities are $E_1 = E_2 = 0.13$. The frequency ratio is $\eta=4$ and $\omega=0.0285\pi$. No clear plateau structure is visible for these laser parameters. }
\label{figu1}       % Give a unique label
\end{figure}

\begin{figure}
\sidecaption
\hspace*{0.10\textwidth}\resizebox{0.30\textwidth}{!}{%
  \includegraphics{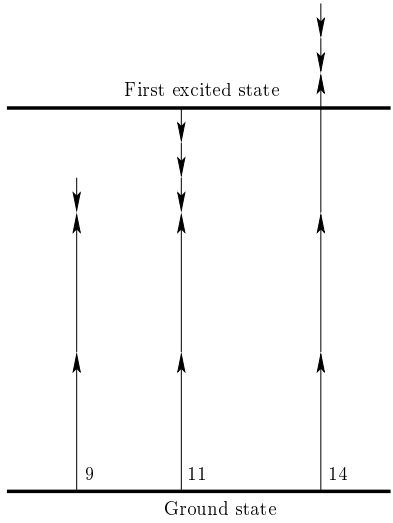}
}
\caption{Energy scheme for the generation of harmonics no.\,9, no.\,11 and no.\,14. The fundamental laser frequency is chosen so that the absorption process leading to the generation of harmonic no. 11 is resonant with the transition between the ground state and the field-free first excited state.}
\label{figu2}      
\end{figure}

\begin{figure}
\sidecaption
\resizebox{0.5\textwidth}{!}{%
  \includegraphics{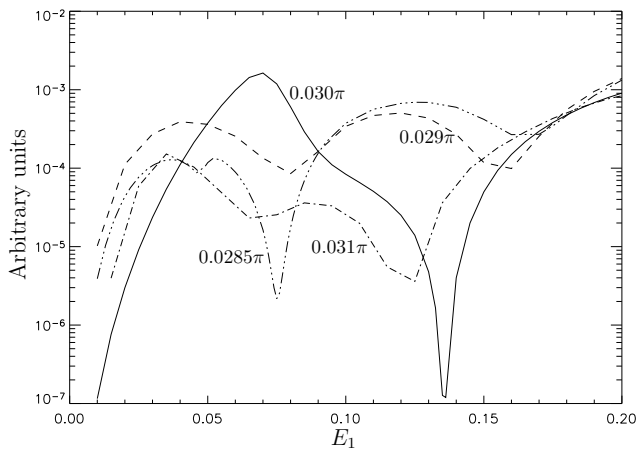}
}
\caption{Dipole strength of harmonic no.\,11 vs $E_1$ for different laser frequencies. $E_2=0.13$ is held constant. A small increase (or decrease) of the laser frequency affects strongly the efficiency conversion of the harmonic which approaches the resonance.}
\label{figu5}      
\end{figure}

\begin{figure}
\sidecaption
\resizebox{0.5\textwidth}{!}{%
  \includegraphics{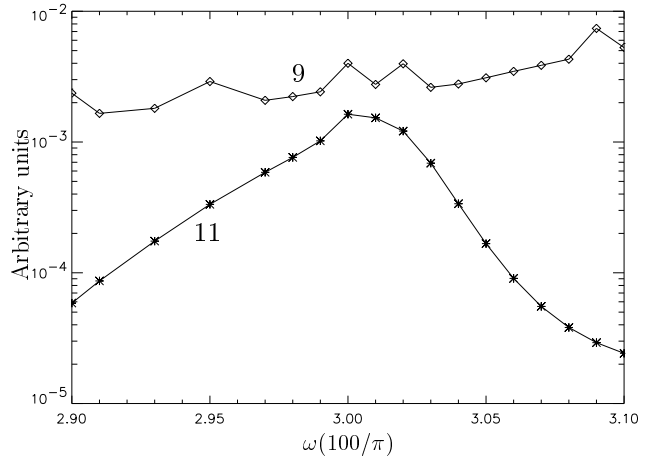}
}
\caption{Dipole strength of harmonics no.\,9 and no.\,11 vs laser frequency. A strong enhancement of harmonic no.\,11  due to the resonance is observed. }
\label{figu6}      
\end{figure}

The numerical simulations were performed through an {\em ab initio} integration of the time-dependent Schr\"odinger equation on a two-dimensional (2D) grid. Choosing a 2D grid permits to run simulations quickly on every modern PC and is not expected to introduce qualitative modifications to the phenomena we aim to investigate. In particular, this holds for the specific configuration we are addressing here. In fact, both the electric field vectors of the lasers as well as the electron's trajectories lie in the plane that is represented by the 2D numerical grid. In the direction perpendicular to that plane the electron wave packet merely spreads. This spreading is not expected  to strongly affect the main features of the harmonic generation spectra such as the cut-off, for instance. Using a polar grid rather than a Cartesian one ensures that no artificial symmetry violation due to the numerics is introduced.\\
In polar coordinates ($\rho, \varphi$), dipole approximation, and length gauge the time-dependent Schr\"odinger equation of our system reads
\begin{eqnarray}
{\rm i}\frac{\partial}{\partial t} \Psi(\rho,\varphi,t) &&= \Bigg [ -\frac{1}{2\rho}\frac{\partial}{\partial \rho} - \frac{1}{2\rho^2}\frac{\partial^2}{\partial \varphi^2} - \frac{\partial^2}{\partial z^2} + V_{\mbox{\scriptsize at}}(\rho) \nonumber \\
&&+ \sin^2(\Theta t) \Big (E_1 \rho \, \mbox{cos}(\varphi -\omega t) \nonumber \\ 
&&+ E_2 \rho \, \mbox{cos}\,(\varphi + \eta \omega t) \Big ) \Bigg ] \Psi(\rho,\varphi,t)
\end{eqnarray}
where the two laser pulses have a duration $T=\pi/\Theta$, with $\Theta = \omega/84$,  and a sine-square shape.  $V_{\mbox{\scriptsize at}}(\rho)$ is a ``soft-core'' 2D potential given by\begin{equation}
V_{\mbox{\scriptsize at}}(\rho) = -\frac{\alpha}{\sqrt{\rho^2 + \beta^2}}.
\end{equation}
The parameters $\alpha$ and $\beta$ can be tuned in order to adjust the ionization energy. In our simulations we used $\alpha = 2.44$ and $\beta = 0.20$. These values provide an ionization potential of $I_p = 2.0$, i.e., the one of real He${}^+$. 
As we aim to address the role played by resonances in the conversion efficiency it is important to know the level scheme of the model potential we use. With the chosen parameters the lowest four excited states have energies $\Omega_1 = 0.985$, $\Omega_2 = 1.375$, $\Omega_3 = 1.548$, and $\Omega_4 = 1.592$. These energies are measured with respect to the ground state. The laser frequencies have been chosen in accordance to this levels scheme. Concerning the frequency ratio $\eta$ all the results presented hereafter have been obtained taking $\eta = 4$. In Fig.\,\ref{figu1} an example of a two-color spectrum obtained with laser fields of equal intensities is shown. As expected only orders allowed by the selection rules are present. Substructures and broadening of certain harmonics can be present when bound states other than the ground state come into play\cite{cecch1}. \\
The laser frequency $\omega$ has to be chosen carefully. In fact, on one hand we want the lower harmonics not to be affected by the atomic levels, on the other hand we want to approach resonances for the higher harmonics. 
The total absorbed energy which leads to harmonic generation is $\omega N^*$ with $N^* = k + \eta(k \pm 1)$. Obviously $N^*$ is also the order of the emitted harmonic. Therefore choosing $\omega = \Omega_i/N^*$ provides the ``virtual'' resonance, i.e., the $N^*$-th harmonic is resonant with the transition between the ground state and the $i$-th excited state. This is a particular kind of resonance, in fact none of the two absorption processes is directly resonant, but their combination is. The ionization rate is nearly not affected by the closeness of such kind of resonances and remains always very low. Choosing $N^* = 11$ and $i = 1$ the expected resonant frequency is $\omega = 0.0285\pi$. A scheme of such a configuration is presented in Fig.\,\ref{figu2}. While the harmonic no. 11 is exactly resonant with the first excited state, harmonics no. 6 and no. 9 are not affected by any energy level. 
\begin{figure}
\sidecaption
\resizebox{0.5\textwidth}{!}{%
  \includegraphics{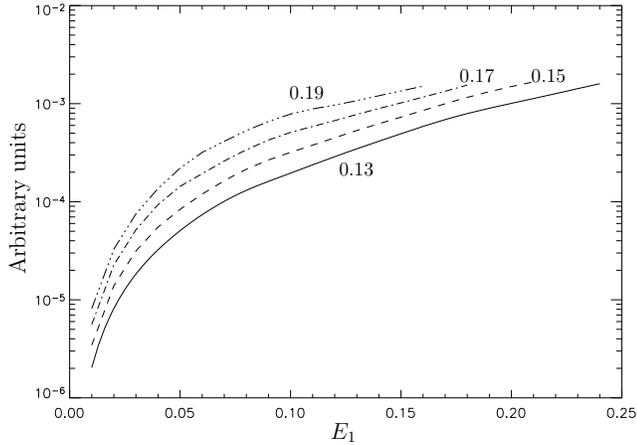}
}
\caption{Dipole strength for the harmonic no.\,6 for different values of $E_2$. The laser frequency is $\omega=0.0285\pi$. The behavior is purely perturbative. }
\label{figu3}      
\end{figure}
Once the frequency value was fixed, a series of simulations for different laser fields intensities have been performed. As we deal with two different absorption processes, in order to obtain efficient harmonic generation it is necessary that the two separate processes are ``likely'',i.e., the probabilties of absorbing the required numbers of photons from each of the two laser must be of the same order of magnitude. \\
\begin{figure}
\sidecaption
\resizebox{0.5\textwidth}{!}{%
  \includegraphics{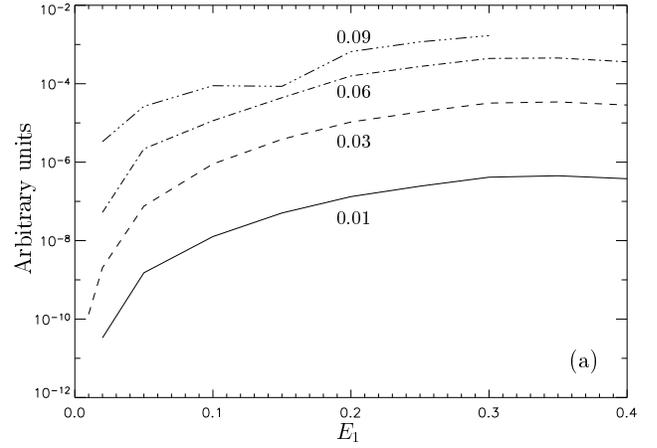}
}
\resizebox{0.5\textwidth}{!}{%
  \includegraphics{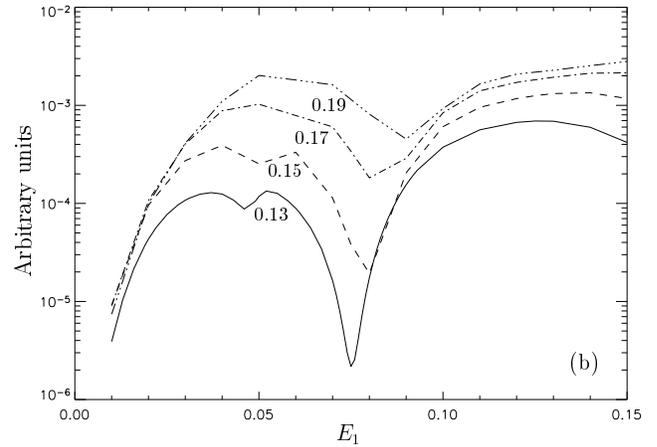}
}
\caption{Dipole strength for the harmonic no.\,11 for low (a) and high (b) values of $E_2$. The laser frequency is $\omega=0.0285\pi$. A perturbative behavior is observed only in the case of relatively weak fields.  }
\label{figu4}      
\end{figure}
In Fig.\,\ref{figu3} the behavior of the harmonic no. 6 versus the electric field $E_1$ for different values of $E_2$ is plotted. As expected the intensity of the harmonic increases smoothly with increasing laser intensity. A very different behavior is observed for the harmonic no. 11 (Fig.\,\ref{figu4}). For relatively low intensities of $E_1$ the behavior is similar to that of the harmonic no. 6, but when the electric field $E_1$ becomes stronger a much more complex behavior appears. This is clearly due to the resonance with the first excited state. Actually, the frequency $\omega = \Omega_1/11$ is close to resonance, but not exactly resonant. In fact, one should take into account the level shifts due to the dynamical Stark effect: all levels are expected to be moved upward. Giving an analytical estimation on how much the lower lying excited states are moved up is a very difficult task already when only a single, linearly polarized laser field is taken into account, let alone in our two-color configuration. In particular it is difficult to distinguish between the different contributions of the two laser fields.\\
A possible method for studying how a full resonance affects the harmonic generation is to perform a series of numerical simulations where all parameters but the laser frequency $\omega$ is kept constant. Increasing the energy of the photons compensates the shift due to the Stark effect and allows to reach the shifted excited state again with 11 photons. As the amount of the shift is unkwown different values of $\omega$ have been used. The results of these simulations are presented in Fig.\,\ref{figu5}. It appears that the harmonic intensity strongly depends on the exact value of the laser frequency, i.e., a small increase or decrease in the frequency value can change the harmonic intensity several orders of magnitude. From Fig.\,\ref{figu5} one realizes that the full resonance is achieved for $\omega = 0.030\pi$ and $E_1 = 0.07$, leading to an energy shift which is about 0.052 a.u. All these values are obtained keeping constant $E_2= 0.13$. If the value of $E_2$ is different the previous values for $\omega$ and $E_1$  do not hold anymore although the physics remains qualitatively the same.
In order to better estimate the dependence of the conversion efficiency with respect to the exact location of the excited state a different kind of study has been carried out. Taking the laser intensities which in Fig.\,\ref{figu5} give the maximum harmonic intensity, namely $E_1 = 0.07$ and $E_2 = 0.13$, a series of simulations for different laser frequencies within a small frequency range has been performed. The results of such study are presented in Fig.\,\ref{figu6}. While the intensity of harmonic no.\,9 varies slowly, the intensity of harmonic no. 11 exhibits a strong enhancement due to the resonance. The ionization level remains negligible for all frequencies values presented in Fig.\,\ref{figu6}. 
\section*{Conclusions}
In this work an extensive numerical study of harmonic generation in the two-color coplanar configuration has been presented. In particular the cases of far-off and near resonances absorption have been addressed. It has been shown that the atomic levels can be used as an important tool in order to enhance  significantly the intensity of a particular harmonic without increasing  the ionization rate. 
\section*{Acknowledgment}
This work was supported in part by INFM through the Advanced Research Project CLUSTERS and in part by DFG. The possibility of using the calculation facility at PC\,${}^2$ in Paderborn, Germany, is  gratefully acknowledged.

%
% BibTeX users please use
% \bibliographystyle{}
% \bibliography{}

\end{document}